\documentclass[11pt]{amsart}
\usepackage{amsmath,amstext,amsthm,amsfonts,amssymb,amsthm}
\usepackage{multicol}
\usepackage{latexsym}
\usepackage{color}
\usepackage{graphicx}
\usepackage{epsf}
\usepackage{epsfig}
\usepackage{subfigure}
\usepackage{rotating}
\usepackage{curves}
\usepackage{epic}

\usepackage{graphics}
\usepackage{verbatim}
\usepackage{color}
\usepackage{hyperref}

\newtheorem{theorem}{Theorem}[section]

\theoremstyle{definition}

\theoremstyle{remark}

\numberwithin{equation}{section}
\newcommand{\abs}[1]{\lvert#1\rvert}

\newcommand{\RB}{\mathbb{R}}

\newcommand{\C}{\mathbb{C}}
\newcommand{\HQ}{\mathbb{H}}

\usepackage{amssymb}
\usepackage{amstext}
\usepackage{amsmath}
\usepackage{latexsym}
\usepackage{color}

\newcommand{\quat}{\mathbb H}

\newcommand{\be}{\begin{equation}}
\newcommand{\en}{\end{equation}}

\newcommand{\bedefin}{\begin{defi}}
\newcommand{\findefi}{\end{defi} \medskip}

\newcommand{\betheo}{\begin{theorem}$\!\!${\bf \,\,\,}}
\newcommand{\entheo}{\end{theorem}}
\newcommand{\enth}{\end{theorem}}

\newcommand{\becor}{\begin{cor}$\!\!${\bf .}}
\newcommand{\encor}{\end{cor}}

\newcommand{\belem}{\begin{lem}$\!\!${\bf .}}
\newcommand{\enlem}{\end{lem}}

\newcommand{\bea}{\begin{eqnarray}}
\newcommand{\ena}{\end{eqnarray}}

\newcommand{\beano}{\begin{eqnarray*}}
\newcommand{\enano}{\end{eqnarray*}}

\newcommand{\bee}{\begin{enumerate}}
\newcommand{\ene}{\end{enumerate}}

\newcommand{\bei}{\begin{itemize}}
\newcommand{\eni}{\end{itemize}}

\newcommand{\betab}{\begin{tabular}}
\newcommand{\entab}{\end{tabular}}

\newcommand{\bd}{\begin{displaymath}}

\newcommand{\h}{{\mathfrak H}}
\newcommand{\hquat}{\mbox{\boldmath $\mathfrak H_{\mathbb H}$}}
\newcommand{\kc}{{\mathfrak K_{\mathbb C}}}

\newcommand{\bsigma}{\mbox{\boldmath $\sigma$}}

\newcommand{\bfeta}{\mbox{\boldmath $\eta$}}

\newcommand{\bPhi}{\mbox{\boldmath $\Phi$}}

\newcommand{\bbra}{\mbox{\boldmath $($}}
\newcommand{\bket}{\mbox{\boldmath $)$}}
\newcommand{\bmid}{\mbox{\boldmath $\;\mid\;$}}

\newcommand{\bfrakf}{\mbox{\boldmath $\mathfrak f$}}

\newcommand{\bfrakq}{\mbox{\boldmath $\mathfrak q$}}

\newcommand{\bfrakx}{\mbox{\boldmath $\mathfrak x$}}
\newcommand{\bfraka}{\mbox{\boldmath $\mathfrak a$}}
\newcommand{\bfrakb}{\mbox{\boldmath $\mathfrak b$}}
\newcommand{\bfrako}{\mbox{\boldmath $\mathfrak o$}}
\newcommand{\bfrakk}{\mbox{\boldmath $\mathfrak k$}}

\newcommand{\bA}{\mathbf A}
\newcommand{\ba}{\mathbf a}
\newcommand{\bb}{\mathbf b}
\newcommand{\bK}{\mathbf K}
\newcommand{\bk}{\mathbf k}

\newcommand{\bq}{\mathbf q}

\newcommand{\bx}{\mathbf x}

\newcommand{\bU}{\mathbf U}
\newcommand{\bi}{\mathbf i}
\newcommand{\bj}{\mathbf j}

\newcommand{\qu}{\mathbf{q}}

\begin{document}
\title[Quaternionic wavelets ]{The Quaternionic Affine Group and Related Continuous Wavelet Transforms on Complex and Quaternionic Hilbert Spaces}
\author{S. Twareque Ali$^{1}$}
\address{$^1$Department of Mathematics and
Statistics, Concordia University,
Montr\'eal, Qu\'ebec, H3G 1M8, Canada.}
\author{K. Thirulogasanthar$^{2}$}
\address{$^{2}$ Department of Computer Science and Software
Engineering, Concordia University, 1455 De Maisonneuve Blvd. West,
Montr\'eal, Qu\'ebec, H3G 1M8, Canada. }
\email{twareque.ali@concordia.ca}
\email{santhar@gmail.com}

\thanks{The research of STA was partly supported by the Natural Sciences and Engineering Research Council of Canada (NSERC)}
\subjclass{Primary
81R30, 81R10, 22E45}
\date{\today}
\keywords{Affine group, square-integrable representation, Quaternion, Wavelets}
\begin{abstract}
  By analogy with the real and complex affine groups, whose unitary irreducible representations are used to define the one and two-dimensional continuous wavelet transforms, we study here the quaternionic affine group and construct its unitary irreducible representations. These representations are constructed both on a complex and a quaternionic Hilbert space. As in the real and complex cases, the representations for the quaternionic group also turn out to be square-integrable.   Using these representations we constrct quaternionic wavelets and continuous wavelet transforms on both the complex and quaternionic Hilbert spaces.
\end{abstract}

\maketitle
\pagestyle{myheadings}

\section{Introduction}\label{sec_intro}
The continuous wavelet transform (CWT), as used extensively in signal analysis and image processing, is a joint time frequency transform. This is in sharp contrast to the Fourier transform, which can be used either to analyze the frequency content of a signal, or its time profile, but not both at the same time.  In the real case, the CWT is built out of the coherent states obtained from a unitary, irreducible and square-integrable representation of the one-dimensional affine group, $\mathbb R \rtimes \mathbb R^*$, a group of translations and dilations of the real line.

In actual practice, for computational purposes, one uses a discretized version of the transform. But the advantage of working with the CWT is that starting with it, one can obtain many more than one discrete transform. In the real case, a signal can be identified with an element $f\in L^2(\mathbb{R}, dx)$ and its norm, $\|f\|^2$, is identified with the energy of the signal. The wavelet transform is built out of a single element $\psi\in L^2(\mathbb{R}, dx)$ and it has to be an admissible vector in the sense of square-integrable representations. In the signal analysis literature such a vector is called a mother wavelet. Then the resulting resolution of the identity enables one to reconstruct the signal from its wavelet transform. For detailed description one could refer, for example, to \cite{Ali,Van,Dau}.

For the complex case, exactly as in the one dimensional situation, wavelets can be derived from the complex affine group, $\mathbb C\rtimes \mathbb C^*$, which is also known as the similitude group of $\mathbb{R}^2$, denoted  $SIM(2)$, consisting of dialations, rotations and translations of the plane. From the point of view of  applications, these wavelets have became  standard tools, for example, in image processing including radar imaging. For details see \cite{Ali} and the many references therein.

 Recently a number of papers have appeared which construct quaternion valued continuous wavelet transforms \cite{Maw1, Maw2,Zha}.  In  \cite{Maw1, Maw2} the authors have contructed a CWT from the group $SIM(2)$ on the Hilbert space $L^2(\RB^2, \quat)$ of quaternion valued functions ($\quat$ is the field of quaternions) and obtained a  reconstruction formula. In \cite{Zha} the admissibility criterion for a CWT of the group $IG(2)=(\RB^+\times SO(2))\otimes\RB^2$ on the Hilbert space $L^2(\RB^2, \quat)$ has been studied. To the best of our knowledge, a CWT from the quaternionic affine group, $\mathbb H\rtimes\mathbb H^*$, either on a complex or on a quaternioninc Hilbert space has not been attempted. In this regard, the novelty of the present work can be stated as follows: following the same procedure as in the real and complex cases, by appropriately identifying a unitary irreducible and square-integrable representation (UIR) of the quaternionic affine group, $\mathbb H\rtimes\mathbb H^*$, in a complex Hilbert space and in a quaternion Hilbert space, we construct two classes of continuous wavelet transforms, study their square-integrability properties and obtain reconstruction formulae. A continuous wavelet transform on a quaternionic Hilbert space involves two complex functions. This means, compared to a wavelet transform on a complex Hilbert space, we get here two transforms. Moreover, the quaternionic affine group, which we later also call the dihedral similitude group of $R^4$,  is similar, though not quite, to having two copies of $SIM(2)$. The hope is that such a transform could be useful in studying stereophonic or stereoscopic signals.

We systematically use a $2\times 2$ complex matrix representation for the quaternions (as is common in the physical literature). This turns the algebraic manipulations, both at the quaternionic and Hilbert space levels, into matrix multiplications, and also renders the form of the quaternionic Hilbert space elements more familiar to physicists and signal analysts. Since the use of quaternions and quaternionic Hilbert spaces may not be familiar to many readers, we recapitulate the notation, convention and some elementary properties in a bit of detail. Also, in order to draw the parallel between the quaternionic affine group and the real and complex affine groups we discuss in some detail these latter two groups also and recapitulate some known facts about their related wavelet transforms.

The article is organized as follows. In Section 2, we list some properties of quaternions, introduce the $2\times 2$-complex matrix representation and give the definitions of quaternionic Hilbert spaces. In Section 3 we briefly recall the one and two dimensional continuous wavelet transform. In Section 4 we discuss the quaternionic affine group, its Haar measures, and study the wavelet transform on a complex Hilbert space. Section 5 deals with the continuous wavelet transform of the quaternionic affine group  on a quaternionic Hilbert space. We conclude with some discussion and indication of future lines of work.

\section{Mathematical preleminaries}
In order to make the paper self-contained, we recall a few facts about quaternions which may not be well-known. In particular, we revisit the $2\times 2$ complex matrix representations of quaternions and quaternionic Hilbert spaces. For details one could refer \cite{Ad, Gra1}.

\subsection{Quaternions}
Let $\quat$ denote the field of all quaternions and $\quat^*$ the group (under quaternionic
multiplication) of all
invertible quaternions. A general quaternion can be written as
$$\bfrakq = q_0 + q_1 \bi + q_2 \bj + q_3 \bk, \qquad q_0 , q_1, q_2, q_3 \in \mathbb R, $$
where $\bi,\bj,\bk$ are the three quaternionic imaginary units, satisfying
$\bi^2 = \bj^2 = \bk^2 = -1$ and $\bi\bj = \bk = -\bj\bi,  \; \bj\bk = \bi = -\bk\bj,
   \; \bk\bi = \bj = - \bi\bk$. The quaternionic conjugate of $\bfrakq$ is
$$ \overline{\bfrakq} = q_0 - \bi q_1 - \bj q_2 - \bk q_3 . $$
We shall use the
$2\times 2$ matrix representation of the quaternions, in which
$$
   \bi = \sqrt{-1}\sigma_1, \quad \bj = -\sqrt{-1}\sigma_2, \quad \bk  = \sqrt{-1}\sigma_3, $$
and the $\sigma$'s are the three Pauli matrices,
$$
  \sigma_1 = \begin{pmatrix} 0 & 1\\ 1& 0 \end{pmatrix}, \quad
\sigma_2 = \begin{pmatrix} 0 & -i\\ i & 0\end{pmatrix}, \quad
\sigma_3 = \begin{pmatrix} 1 & 0\\ 0 & -1 \end{pmatrix}, $$
to which we add
$$\sigma_0 = \mathbb I_2 = \begin{pmatrix} 1 & 0 \\ 0 & 1\end{pmatrix}.$$
We shall also use the matrix valued vector $\bsigma = (\sigma_1, -\sigma_2 , \sigma_3)$. Thus,
in this representation,
$$\bfrakq = q_0\sigma_0 + i\bq\cdot \bsigma =
\begin{pmatrix} q_0 + iq_3 & -q_2 +iq_1 \\ q_2 + iq_1 & q_0 -1q_3\end{pmatrix}, \quad
\bq = (q_1, q_2, q_3 ).  $$
In this representation, the quaternionic conjugate of $\bfrakq$ is given by $\bfrakq^\dag$.
Introducing  two complex variables, which we write as
$$ z_1 = q_0 + iq_3 , \qquad z_2 = q_2 + iq_1, $$
we may also write
\be
  \bfrakq = \begin{pmatrix} z_1 & -\overline{z}_2\\ z_2 & \overline{z}_1 \end{pmatrix}.
\label{comp-rep-quat}
\en
From this it is clear that the group $\quat^*$ is isomormphic to the {\em affine $SU(2)$ group},
i.e., $\mathbb R^{>0}\times SU(2)$, which is the group $SU(2)$ together with all (non-zero)
dilations. As a set  $\quat^* \simeq \mathbb R^{>0} \times S(4)$, where $S(4)$ is the surface of the sphere, or more simply, $\quat^* \simeq \mathbb R^4 \backslash \{\mathbf 0\}$.
From (\ref{comp-rep-quat}) we get
\be
  \text{det}[\bfrakq] = \vert z_1\vert^2 + \vert z_2\vert^2 = q_0^2 + q_1^2 + q_2^2 + q_3^2
     : = \vert \bfrakq \vert^2 ,
\label{quat-norm}
\en
$\vert \bfrakq \vert$ denoting the usual norm of the quaternion $\bfrakq$. Note also that
$$
  \bfrakq^\dag \bfrakq =   \bfrakq \bfrakq^\dag = \vert \bfrakq \vert^2\; \mathbb I_2 . $$
If $\bfrakq$ is invertible,
$$
  \bfrakq^{-1} =  \frac 1{\vert \bfrakq \vert^2 }
  \begin{pmatrix} \overline{z}_1  &  \overline{z}_2 \\ -z_2 & z_1 \end{pmatrix} .$$

\subsection{Quaternionic Hilbert spaces}
In this subsection we  define left and right quaternionic Hilbert spaces. For details we refer the reader to \cite{Ad, Gra1}. We also define the Hilbert space of square-integrable functions on quaternions based on \cite{Vis}.
\subsubsection{Right Quaternionic Hilbert Space}
Let $V_{\quat}^{R}$ be a linear vector space under right multiplication by quaternions.  For $f,g,h\in V_{\quat}^{R}$ and $\bfrakq\in \quat$, the inner product
$$\langle\cdot\mid\cdot\rangle:V_{\quat}^{R}\times V_{\quat}^{R}\longrightarrow \quat$$
satisfies the following properties
\begin{enumerate}
\item[(i)]
$\overline{\langle f\mid g\rangle}=\langle g\mid f\rangle$
\item[(ii)]
$\|f\|^{2}=\langle f\mid f\rangle>0$ unless $f=0$, a real norm
\item[(iii)]
$\langle f\mid g+h\rangle=\langle f\mid g\rangle+\langle f\mid h\rangle$
\item[(iv)]
$\langle f\mid g\bfrakq\rangle=\langle f\mid g\rangle\bfrakq$
\item[(v)]
$\langle f\bfrakq\mid g\rangle=\overline{\bfrakq}\langle f\mid g\rangle$
\end{enumerate}
where $\overline{\bfrakq}$ stands for the quaternionic conjugate. It is always assumed that the
space $V_{\quat}^{R}$ is complete under the norm given above. Then,  together with $\langle\cdot\mid\cdot\rangle$ this defines a right quaternionic Hilbert space. Quaternionic Hilbert spaces share most of the standard properties of complex Hilbert spaces. The Dirac bra-ket notation
can be adapted to quaternionic Hilbert spaces:
$$\mid f\bfrakq\rangle=\mid f\rangle\bfrakq,\hspace{1cm}\langle f\bfrakq\mid=\overline{\bfrakq}\langle f\mid\;, $$
for a right quaternionic Hilbert space, with $\vert f\rangle$ denoting the vector $f$ and $\langle f\vert$ its dual vector.

\subsubsection{Left Quaternionic Hilbert Space}
Let $V_{\quat}^{L}$ be a linear vector space under left multiplication by quaternions.  For $f,g,h\in V_{\quat}^{L}$ and $\bfrakq\in \quat$, the inner product
$$\langle\cdot\mid\cdot\rangle:V_{\quat}^{L}\times V_{\quat}^{L}\longrightarrow \quat$$
satisfies the following properties
\begin{enumerate}
\item[(i)]
$\overline{\langle f\mid g\rangle}=\langle g\mid f\rangle$
\item[(ii)]
$\|f\|^{2}=\langle f\mid f\rangle>0$ unless $f=0$, a real norm
\item[(iii)]
$\langle f\mid g+h\rangle=\langle f\mid g\rangle+\langle f\mid h\rangle$
\item[(iv)]
$\langle \bfrakq f\mid g\rangle=\bfrakq\langle f\mid g\rangle$
\item[(v)]
$\langle f\mid \bfrakq g\rangle=\langle f\mid g\rangle\overline{\bfrakq}$
\end{enumerate}
Again, we shall assume that the space $V_{\quat}^{L}$ together with $\langle\cdot\mid\cdot\rangle$ is a separable Hilbert space. Also,
\begin{equation}\label{leftcs}
\mid \bfrakq f\rangle=\mid f\rangle\overline{\bfrakq},\hspace{1cm}\langle \bfrakq f\mid=\bfrakq\langle f\mid.
\end{equation}
Note that, because of our convention for inner products, for a left quaternionic Hilbert space, the bra vector $\langle f\mid$ is to be identified with the vector itself, while the ket vector $\mid f \rangle$ is to be identified with its dual.
(There is a natural left multiplication by quaternionic scalars on the dual of a right quaternionic Hilbert space and a similar right multiplication on the dual of a left quaternionic Hilbert space.)

The field of quaternions $\quat$ itself can be turned into a left quaternionic Hilbert space by defining the inner product $\langle \bfrakq \mid \bfrakq^\prime \rangle = \bfrakq \bfrakq^{\prime\dag} = \bfrakq\overline{\bfrakq^\prime}$ or into a right quaternionic Hilbert space with  $\langle \qu \mid \qu^\prime \rangle = \bfrakq^\dag \bfrakq^\prime = \overline{\bfrakq}\bfrakq^\prime$.
\subsubsection{Quaternionic Hilbert Spaces of Square-integrable Functions}
Let $(X, \mu)$ be a measure space and $\quat$  the field of quaternions, then
$$L^2_{\quat}(X,\mu)=\left\{f:X\rightarrow \quat\;\; \left| \;\; \int_X|f(x)|^2d\mu(x)<\infty \right.\right\}$$\label{L^2}
is a left quaternionic Hilbert space, with the (left) scalar product
\begin{equation}
\langle f \mid g\rangle =\int_X f(x)\overline{g(x)} d\mu(x),
\label{left-sc-prod}
\end{equation}
where $\overline{g(x)}$ is the quaternionic conjugate of $g(x)$, and (left)  scalar multiplication $\bfraka f, \; \bfraka\in \quat,$ with $(\bfraka f)(x) = \bfraka f(x)$ (see \cite{Vis} for details). Similarly, one could define a right quaternionic Hilbert space of square-integrable functions.
\section{CWT on real and complex affine groups}
Since the construction of CWT for the quaternionic affine group follows its real and complex counterparts, and for the sake of completeness, in this section we shall revisit the CWT of real and complex affine groups very briefly.

\subsection{The real case}
Let $\mathbb{R}^*=\{a\in\mathbb{R}~|~a\not=0\}$ and consider the semi-direct product group
$$G^{\mathbb{R}}_{\text{aff}}=\mathbb{R}\rtimes\mathbb{R}^*=\{(b,a)~|b\in\mathbb{R}, a\not=0\}$$
with the group operation
$$(b,a)(b',a')=(b+ab',aa')$$
or equivalently
$$G^{\mathbb{R}}_{\text{aff}}=\left\{g=\left(\begin{array}{cc}a&b\\0&1\end{array}\right)\vert b\in\mathbb{R}, a\not=0\right\}$$
with composition given by matrix multiplication,
$$gg'=\left(\begin{array}{cc}a&b\\0&1\end{array}\right)\left(\begin{array}{cc}a'&b'\\0&1\end{array}\right)
=\left(\begin{array}{cc}aa'&b+ab'\\0&1\end{array}\right).$$
An affine transformation on $\mathbb R$ is given by the action of the group $G^{\mathbb{R}}_{\text{aff}}$:
$$x\mapsto ax+b;\quad x\in\mathbb{R}.$$
On $L^2(\mathbb{R},dx)$ define $U:G^{\mathbb{R}}_{\text{aff}}\longrightarrow L^2(\mathbb{R},dx)$ by
$$\left[U(a,b)f\right](x)=\frac{1}{\sqrt{\abs{a}}}f\left(\frac{x-b}{a}\right).$$
Then it is well-known that the representation $U$ is unitary, irreducible and square-integrable with respect to the left Haar measure $\displaystyle d\mu_{l}(b,a)=\frac{dbda}{\abs{a}^2}$. The square-integrability of the representation $U(b,a)$ means that there exist vectors $\psi\in L^2(\RB, dx)$ for which the matrix element $\langle U(b,a)\psi\vert\psi\rangle$ is square-integrable as a function of $b,a$ with respect to the Haar measure,
$$\int\int_{G^{\RB}_{\text{aff}}}d\mu_l(b,a)\abs{\langle U(b,a)\psi\vert\psi\rangle}^2<\infty.
$$
Also, it is a fact that the existence of one such nonzero vector, $\psi$, implies the existence of an entire dense subset of them. Indeed, the condition for a vector to be of this type is precisely the condition of admissibility required of mother wavelets.

In terms of Fourier transforms, the vector $\psi\in L^2(\RB, dx)$ is admissible if and only if
$$c_{\psi}=2\pi\int_{-\infty}^{\infty}\frac{d\zeta}{\abs{\zeta}}\abs{\hat{\psi}}^2<\infty,$$
where $\hat{\psi}$ is the Fourier transform of $\psi$. Defining an operator $\hat{C}$ on $L^2(\hat{\RB},d\zeta)$ as
$$(\hat{C}\hat{\psi})(\zeta)=\left[\frac{2\pi}{\abs{\zeta}}\right]^{1/2}\hat{\psi}(\zeta)$$
and denoting by $C$ its inverse Fourier transform, we see that the vector $\psi$ is admissible if and only if
$$c_{\psi}=\|C\psi\|^2<\infty.$$
The operator $C$ is known as the {\em Duflo-Moore operator}. Let $\psi$ be an admissible vector and $s$ be a signal, i.e., an element of $L^2(\RB, dx)$. Then the wavelet transform of the signal $s$ is
$$S(b,a)=\langle\psi_{b,a}\vert s\rangle$$
and we require that the energy of the transformed signal be finite, i.e., 
$$E(s)=\int\int_{G^{\RB}_{\text{aff}}}d\mu_l(b,a)\abs{S(b,a)}^2<\infty.$$
From this one gets the resolution of the identity,
$$\frac{1}{c_{\psi}}\int\int_{G^{\RB}_{\text{aff}}}d\mu_l(b,a)\vert\psi_{b,a}\rangle\langle\psi_{b,a}\vert=I,$$
which then leads to the celebrated reconstruction formula of signal analysis,
$$s(x)=\frac{1}{c_{\psi}}\int\int_{G^{\RB}_{\text{aff}}}d\mu_l(b,a)S(b,a)\psi_{b,a}(x)\qquad \text{a.e.}.$$


\subsection{The complex case}
Let $\C^*=\{z\in\C~\vert~z\not=0\}$ and $G^{\C}_{\text{aff}}=\C\rtimes\C^*$ with  group operation
$$(z,w)(z',w')=(z+wz',ww').$$
Let $\underline{b}=(b_1,b_2)^T$,  $r_{\theta}=\left(\begin{array}{cc}\cos\theta&\sin\theta\\-\sin\theta&\cos\theta\end{array}\right)\in SO(2);~~\theta\in[0,2\pi)$ and $\lambda>0$, then $G^{\C}_{\text{aff}}$ can be considered as
$$G^{\C}_{\text{aff}}=SIM(2)=\left\{(\underline{b},\lambda,r_{\theta})~\vert~ \underline{b}\in\mathbb{R}^2, \lambda>0, r_{\theta}\in SO(2)\right\}.$$
The affine action can be written as
$$\underline{x}\mapsto \lambda r_{\theta}\underline{x}+\underline{b};\qquad\underline{x}\in\mathbb{R}^2\approxeq\C.$$
Note that $G^{\C}_{\text{aff}}$ can also be represented in matrix form:
$$G^{\C}_{\text{aff}}=\left\{g=\left(\begin{array}{cc}\lambda r_{\theta}&\underline{b}\\ \underline{0}^T&1\end{array}\right)~\vert~\underline{b}\in\mathbb{R}^2, \lambda>0, r_{\theta}\in SO(2)\right\}.$$
Now consider the group representation $U:G^{\C}_{\text{aff}}\longrightarrow L^2(\mathbb{R}^2)$ by
$$[U(\underline{b},\lambda,r_{\theta})f](\underline{x})
=\frac{1}{\lambda}f\left(\frac{r_{\theta}^{-1}(\underline{x}-\underline{b})}{\lambda}\right)$$
or
$$[U(z,w)f](u)=\frac{1}{|w|}f\left(\frac{u-z}{w}\right).$$
Then, it is known that the representation $U$ is unitary, irreducible and square-integrable with respect to the left Haar measure $\displaystyle d\mu_l(z,w)=\frac{dzdw}{\abs{w}^4}$, $dz, dw$ denoting the Lebesgue measures on the complex plane. 


\section{The quaternionic affine group}
In this section  we discuss the quaternionic affine group, its Haar measures and study its wavelet transform on a complex Hilbert space.
\subsection{Action of $\quat^*$ on $\quat$}\label{subseec-quat-action}
Consider the action of $\quat^*$ on $\quat$ by right (or left) quaternionic (in our
representation matrix) multiplication. It is clear that there are only two orbits
under this action,
$\{\bfrako\}$ (the zero quaternion) and $\quat^*$. Furthermore, this latter orbit is
{\em open and free\/}. Let
$$ \bfraka = \begin{pmatrix} w_1 & -\overline{w}_2\\ w_2 & \overline{w}_1 \end{pmatrix}
\in \quat^* \quad \text{and} \quad
  \bfrakx = \begin{pmatrix} z_1 & -\overline{z}_2\\ z_2 & \overline{z}_1 \end{pmatrix}
  \in \quat .$$
Then under left action
\be \bfrakx \longmapsto \bfrakx^\prime = \bfraka \bfrakx
  = \begin{pmatrix} w_1 z_1 - \overline{w}_2 z_2 & -
  \overline{w}_2\overline{z}_1 -w_1\overline{z}_2 \\ w_2 z_1 + \overline{w}_1 z_2 &
     \overline{w}_1 \overline{z}_1  - w_2 \overline{z}_2\end{pmatrix} .
\label{left-act}
\en
We take $w_1 =a_0 + ia_3, \; w_2 = a_2 + ia_1$ and $z_1 = x_0 + ix_3, \; z_2 =
x_2 + ix_1$ and consider $\bfrakx$ as the vector
\be
  \bx = \begin{pmatrix} x_0 \\x_3 \\ x_2 \\x_1 \end{pmatrix} \in \mathbb R^4 .
\label{vec-rep}
\en
On this vector, the left action (\ref{left-act}) is easily seen to lead to the matrix
left action
\be
  \bx \longmapsto \bx^\prime = A\bx = \begin{pmatrix} a_0 & -a_3 & -a_2 & -a_1 \\
                                                       a_3 & a_0 & a_1 & -a_2 \\
                                                       a_2 & -a_1 & a_0 & a_3 \\
                                                       a_1 & a_2 & -a_3 & a_0 \end{pmatrix}
                                                        \begin{pmatrix} x_0 \\x_3 \\ x_2 \\x_1 \end{pmatrix}
 = \begin{pmatrix} A_1 & -A_2^T \\ A_2 & A_1^T\end{pmatrix}
            \begin{pmatrix} \bx_1 \\ \bx_2 \end{pmatrix},
 \label{mat-left-act}
\en
on $\mathbb R^4$, where
$$
 A_1 = \begin{pmatrix} a_0 & -a_3 \\a_3 & a_0 \end{pmatrix}, \quad
  A_2 = \begin{pmatrix} a_2 & -a_1 \\a_1 & a_2 \end{pmatrix}, \quad
  \bx_1 = \begin{pmatrix} x_0 \\ x_3 \end{pmatrix}, \quad
  \bx_2 = \begin{pmatrix} x_2 \\ x_1 \end{pmatrix}. $$
 The matrices $A_1$ and $A_2$ are rotation-dilation matrices, and may be written in the form
 \be
    A_1 = \lambda_1 \begin{pmatrix} \cos\theta_1 & - \sin\theta_1\\
                                    \sin\theta_1 &  \cos\theta_1 \end{pmatrix}
    = \lambda_1 R(\theta_1), \qquad
    A_2 = \lambda_2 \begin{pmatrix} \cos\theta_2 & - \sin\theta_2 \\
                                    \sin\theta_2 &  \cos\theta_2 \end{pmatrix}
    = \lambda_2 R(\theta_2)
 \label{rot-dil-mat}
 \en
 where
 \be
  \theta_1 = \tan^{-1}\left(\frac {a_3}{a_0}\right), \;
  \theta_2 = \tan^{-1}\left(\frac {a_1}{a_2}\right), \;
    \lambda_1 = \sqrt{a_0^2 + a_3^2}, \;  \lambda_2 = \sqrt{a_1^2 + a_2^2} \; \text{and} \;
  \lambda_1^2 + \lambda_2^2 \neq 0
 \label{rot-dil-cond}
 \en
 and $R(\theta)$ is the $2\times 2$ rotation matrix
 \be
 R(\theta) = \begin{pmatrix} \cos\theta & -\sin\theta \\ \sin\theta & \cos\theta
   \end{pmatrix} .
 \label{rot-mat}
 \en
 Note that
 $$ A^T A = A A^T  = \vert\bfraka\vert^2 \mathbb I_4\; \quad \text{and} \quad
 \text{det}[A] = \vert\bfraka\vert^4 .$$

 From the above it is clear that when $\quat$ is identified with $\mathbb R^4$, the action of $\quat^*$ on $\quat$ is that of two
 two-dimensional rotation-dilation groups (rotations of the two-dimensional plane together
 with radial dilations, where at least one of the dilations is non-zero) acting on $\mathbb R^4$. Consequently, we shall consider elements in $\quat^*$ interchangeably as $2\times 2$ complex
 matrices of the type
 $$
 \bfraka =  \begin{pmatrix} w_1 & -\overline{w}_2\\ w_2 & \overline{w}_1 \end{pmatrix}, \qquad
 \text{det}[\bfraka ] = \vert\bfraka\vert^2 \neq 0 $$
or $4\times 4$ real matrices of the type $A$ in (\ref{mat-left-act}):
\be
 A = \begin{pmatrix} \lambda_1 R(\theta_1) & -\lambda_2 R(-\theta_2)\\
                        \lambda_2 R(\theta_2) & \lambda_1 R(-\theta_1), \end{pmatrix}, \qquad
\text{det}[A] = \vert\bfraka\vert^4  = [\lambda_1^2 + \lambda_2^2]^2\neq 0.
\label{second-mat-rep}
\en
The matrix $A$ has the inverse
$$
  A^{-1} = \frac 1{\lambda_1^2 + \lambda_2^2} \begin{pmatrix} \lambda_1 R(-\theta_1) & \lambda_2 R(-\theta_2)\\
                       - \lambda_2 R(\theta_2) & \lambda_1 R(\theta_1), \end{pmatrix} . $$

\subsection{The quaternionic affine group}\label{sec-quat-aff-grp}
 Let us look at the three affine groups, $G^{\mathbb R}_{\text{aff}}, G^{\mathbb C}_{\text{aff}}$
 and $G^{\mathbb H}_{\text{aff}}$, of the real line, the complex plane and the quaternions,
 respectively. These groups are defined as the semi-direct products
 $$
   G^{\mathbb R}_{\text{aff}} = \mathbb R \rtimes \mathbb R^* , \qquad
   G^{\mathbb C}_{\text{aff}} = \mathbb C \rtimes \mathbb C^* , \qquad
   G^{\mathbb H}_{\text{aff}} = \mathbb H \rtimes \mathbb H^* . $$
Let $\mathbb K$ denote any one of the three fields $\mathbb R, \mathbb C$ or $\mathbb H$ and write $G^{\mathbb K}_{\text{aff}} = \mathbb K \rtimes \mathbb K^*$. A generic element in
$G^{\mathbb K}_{\text{aff}}$ can be written as
$$ g = (b, a) = \begin{pmatrix} a & b \\ 0 & 1 \end{pmatrix}, \quad a \in \mathbb K^* , \;\;
    b \in \mathbb K .$$
Of these, $G^{\mathbb R}_{\text{aff}}$ is the {\em one-dimensional wavelet group} and
$G^{\mathbb C}_{\text{aff}}$, which is isomorphic to the similitude group of the plane (translations, rotations and dilations of the 2-dimensional plane), is the {\em two-dimensional
wavelet group\/.} By analogy we shall call the quaternionic affine group
$G^{\mathbb H}_{\text{aff}}$ the {\em quaternionic wavelet group\/}, which we now analyse in some detail. In the $2\times 2$ matrix representation of the quaternions introduced earlier, we shall
represent an element of $G^{\mathbb H}_{\text{aff}}$ as the $3\times 3$ complex matrix
\be
   g := (\bfrakb, \bfraka) = \begin{pmatrix} \bfraka & \bfrakb \\ \mathbf{0}^T & 1 \end{pmatrix},
    \quad
   \bfraka \in \mathbb H^*, \quad \bfrakb \in \mathbb H, \quad
   \mathbf{0}^T = (0,0) .
\label{quat-aff-mat}
\en
Alternatively, if $A$ is the $4\times 4$ real matrix corresponding to $\bfraka$, through (\ref{left-act}), and $\mathbf b\in \mathbb R^4$ the vector made out of the components $b_0, b_1,
b_2, b_3$ of $\bfrakb$ (see (\ref{vec-rep})),
$$
   \mathbf b = \begin{pmatrix} b_0 \\b_3\\b_2 \\ b_1 \end{pmatrix} , $$
then $g$ may also be written as the $5 \times 5$ real matrix,
\be
   g := (\mathbf b, A) = \begin{pmatrix} A & \mathbf b \\ \mathbf{0}^T & 1 \end{pmatrix},
    \quad \mathbf{0}^T = (0,0,0,0) .
\label{quat-aff-mat-real}
\en
In this real form $G^{\mathbb H}_{\text{aff}}$ may be called the {\em group of dihedral similitude transformations of $\mathbb R^4$}. We shall use both representations of
 $G^{\mathbb H}_{\text{aff}}$  interchangeably.

For each one of these groups $G^{\mathbb K}_{\text{aff}} = \mathbb K \rtimes \mathbb K^*$ there
is exactly one non-trivial orbit of $\mathbb K^*$ in the dual of $\mathbb K$ and this orbit is
open and free. Hence on a complex Hilbert space, each one of these groups has exactly one
irreducible representtion. the irreducible representations of $G^{\mathbb R}_{\text{aff}}$ and
$G^{\mathbb C}_{\text{aff}}$ are well known and have been displayed above.  We compute below the one irreducible representation of $G^{\mathbb H}_{\text{aff}}$, both in a complex and in a quaternionic  Hilbert
space.

\subsection{Invariant measures of $\quat^*$ and $G^{\mathbb H}_{\text{aff}}$}\label{quat-inv-meas}
The group $\quat^*$ is unimodular. To compute the Haar measure, let
$$
 \bfraka = \begin{pmatrix} a_0 + ia_3 & - a_2 + ia_1\\
              a_2 + ia_1 & a_0 - ia_3 \end{pmatrix} , \;\;
  \bfrakx = \begin{pmatrix} x_0 + ix_3 & - x_2 + ix_1\\
              x_2 + ix_1 & x_0 - ix_3 \end{pmatrix} \in \quat^* . $$
Let $\bx \in \mathbb R^4$ be the vector corresponding to $\bfrakx$ (see (\ref{vec-rep}))
and $A$ the matrix of transformation on $\mathbb R^4$ representing the left action
of $\bfraka$ on $\bfrakx$ (see (\ref{mat-left-act})). We write this action as
\be
   \bfrakx \longmapsto \bfrakx^\prime = \bfraka\bfrakx, \quad \Longrightarrow \quad
   \bx \longmapsto \bx^\prime = A\bx .
\label{2-rt-acts}
\en
By (\ref{quat-norm}) $\text{det}[\bfrakx] = \Vert\bx\Vert^2$. Thus,
\be \Vert A\bx\Vert^2 = \text{det}[\bfraka\bfrakx] =
\text{det}[\bfraka]\text{det}[\bfrakx] = \text{det}[A]^{\frac 12}\Vert\bx\Vert^2 ,
\label{det-transf}
\en
the last equality following from (\ref{second-mat-rep}). We define a measure on
$\quat^*$ by
\be
   d\mu_{\quat^*} = \frac {d\bfrakx}{\vert\bfrakx\vert^4} =
   \frac {d\bfrakx}{(\text{det}[\bfrakx])^2} =
   \frac {d\bx}{\Vert \bx \Vert^4}, \quad
   \text{where} \quad  d\bfrakx = d\bx  = dx_0\;dx_3\;dx_2\;dx_1.
\label{inv-meas}
\en
Then in view of (\ref{det-transf}) and the fact that $d\bx^\prime = \text{det}[A]\;d\bx$,
we see that
$$
  \frac {d\bx^\prime}{\Vert \bx^\prime \Vert^4} = \frac {d\bx}{\Vert \bx \Vert^4}, $$
i.e., $d\mu$ is a left invariant measure on $\quat^*$. Similarly, it is also a right
invariant measure.

The group $G^{\mathbb H}_{\text{aff}}$ is non-unimodular. The two invariant measures are  
again easily computed using standard techniques. Let $(\bfrakb_0, \bfraka_0), \;
(\bfrakb, \bfraka) \in G^{\mathbb H}_{\text{aff}}$, or equivalently, we take
$(\bb_0, A_0)$ and $(\bb, A)$ (see (\ref{quat-aff-mat}) -- (\ref{quat-aff-mat-real})).
Then under the left action,
$$ (\bb, A) \longmapsto (\bb', A') = (\bb_0, A_0)(\bb, A)= (\bb_0 + A_0\bb , A_0A) .$$
Using the invariant measure (\ref{inv-meas}) of $\quat^*$ and
the fact that now $d\bb' = \text{det}[A_0]\;d\bb$, we easily see that the measure
\be
  d\mu_\ell (\bb , A) = \frac {d\mathbf b}{\Vert \ba\Vert^4} \; d\mu_{\quat^*} (A)
    = \frac {{d\mathbf b}\;{d\ba}}{\Vert \ba\Vert^8} :=
      \frac {d\bb\; dA}{(\text{det}[A])^2},
\label{left-inv-haar}
\en
which we shall also write as
\be
  d\mu_\ell (\bfrakb , \bfraka) = \frac {d\bfrakb\; d\bfraka}{(\text{det}[\bfraka])^4} .
\label{left-inv-haar2}
\en
Similarly, we find the right Haar measure to be
\be
  d\mu_r (\bb , A) = d\bb \; d\mu_{\quat^*} (A)
    = \frac {{d\mathbf b}\;{d\ba}}{\Vert \ba\Vert^4} :=
      \frac {d\bb\; dA}{\text{det}[A]},
\label{right-inv-haar}
\en
or alternatively written,
\be
  d\mu_r (\bfrakb , \bfraka) = \frac {d\bfrakb\; d\bfraka}{(\text{det}[\bfraka])^2} .
\label{right-inv-haar2}
\en
The {\em modular function} $\Delta$, such that $d\mu_\ell (\bfrakb , \bfraka) =
\Delta (\bfrakb , \bfraka)\; d\mu_r(\bfrakb , \bfraka)$, is
\be
   \Delta (\bfrakb , \bfraka) = \frac 1{(\text{det}[\bfraka])^2}
   = \frac 1{\vert\bfraka\vert^4} = \frac 1{\Vert\ba\Vert^4}
   = \frac 1{\text{det}[A]} := \Delta (\bb , A).
\label{mod-fcn}
\en

\subsection{UIR of $G^{\mathbb H}_{\text{aff}}$ in a complex Hilbert space}
\label{subsec-uir-compl}
From the general theory of semi-direct products of the type $\mathbb R^n \rtimes H$, where
$H$ is a subgroup of $GL(n, \mathbb R)$, and which has open free orbits in the dual of
$\mathbb R^n$, (see. for example, \cite{Ali}, Chapter 9), we know that
$G^{\mathbb H}_{\text{aff}}$ has exactly one unitary irreducible representation
on a complex Hilbert space and
moreover, this representation is square-integrable.  We proceed to construct this
representation (in a Hilbert space over the complexes). Consider the Hilbert space
$\h_{\mathbb C} = L^2_{\mathbb C} (\mathbb R^4, d\bx )$ and on it define the
representation $G^{\mathbb H}_{\text{aff}} \ni (\bb, A) \longmapsto U_{\mathbb C}(\bb , A)$,
\be
   (U_{\mathbb C}(\bb , A) f)(\bx ) = \frac 1{(\text{det}[A])^{\frac 12}}
   f( A^{-1} (\bx - \bb)) ,
   \qquad f \in \h_{\mathbb C} .
\label{comp-quat-rep}
\en
This representation is unitary and irreducible.
  To analyze the square-integrabillity of this representation, we note that from the
general theory \cite{Ali}, the {\em Duflo-Moore operator $C$}  is given in the
Fourier domain as the multiplication operator
\be
   (\widehat{C}\widehat{f})(\bk ) = \mathcal C (\bk )
                     \widehat{f} (\bk), \quad \text{where} \quad
   \mathcal C (\bk ) = \left[\frac {2\pi}{\Vert \bk\Vert}  \right]^2 .
\label{duflo-moore-op}
\en
A vector $f\in \h_{\mathbb C}$ is {\em admissible} if it is in the domain of $C$
i.e., if its Fourier transform $\widehat{f}$ satisfies
$$  (2\pi)^4\int_{\mathbb R^4} \frac {\vert \widehat{f}(\bk )\vert^2}
            {\Vert \bk \Vert^4} \; d\bk < \infty. $$
Thus, for any two vectors $\eta_1, \eta_2$ in the domain of $C$ and for arbitrary
$f_1, f_2 \in \h_{\mathbb C}$, we have the {\em orthogonality relation\/},
\be
  \int_{G^{\mathbb H}_{\text{aff}}} \langle f_1 \mid U_{\mathbb C} (\bb , A)\eta_1
  \rangle\langle\eta_2 \mid  U_{\mathbb C} (\bb , A)^*f_2\rangle \; d\mu_\ell (\bb , A)
  = \langle C\eta_2 \mid C\eta_1 \rangle \langle f_1 \mid f_2 \rangle ,
\label{comp-orthog}
\en
which is the same as the operator equation
\be
  \int_{G^{\mathbb H}_{\text{aff}}}  \vert U_{\mathbb C} (\bb , A)\eta_1
  \rangle\langle\eta_2 \vert  U_{\mathbb C} (\bb , A)^* \; d\mu_\ell (\bb , A)
  = \langle C\eta_2 \mid C\eta_1 \rangle I_{\h_{\mathbb C}} \; .
\label{op-comp-orthog}
\en
If $\langle C\eta_2 \mid C\eta_1 \rangle \neq 0$, we have the {\em resolution of the
identity}
\be
  \frac 1{\langle C\eta_2 \mid C\eta_1 \rangle} \int_{G^{\mathbb H}_{\text{aff}}}
      U_{\mathbb C} (\bb , A)\vert\eta_1 \rangle
      \langle \eta_2 \vert U_{\mathbb C} (\bb , A)^* \;  d\mu_\ell (\bb , A)  =
      I_{\h_{\mathbb C}} \; .
\label{comp-resolid}
\en
Given an admissible vector $\eta$, such that $\Vert C\eta\Vert^2 =1$,  we define
the family of {\em coherent states or wavelets} as
\be
  \mathfrak S_{\mathbb C} = \{\eta_{\bb , A} = U_{\mathbb C}(\bb , A )\eta \mid
     (\bb , A ) \in {G^{\mathbb H}_{\text{aff}}}\} ,
\label{cs-wav}
\en
which then satisfies the resolution of the identity,
\be
   \int_{G^{\mathbb H}_{\text{aff}}}
      \vert\eta_{\bb, A} \rangle
      \langle \eta_{\bb , A} \vert \;  d\mu_\ell (\bb , A)  =
      I_{\h_{\mathbb C}} \; .
\label{comp-wav-resolid}
\en

The above representation could also be realized on the Hilbert space
$\kc = L^2_{\mathbb C}(\quat , d\bfrakx )$ over the quaternions. We simply transcribe
Eqs. (\ref{comp-quat-rep}) -- (\ref{comp-wav-resolid}) into this framework.
Thus, we define the
representation $G^{\mathbb H}_{\text{aff}} \ni (\bfrakb, \bfraka) \longmapsto
U_{\mathbb C}(\bfrakb , \bfraka)$,
\be
   (U_{\mathbb C}(\bfrakb , \bfraka) f)(\bfrakx ) = \frac 1{\text{det}[\bfraka]}
   f( \bfraka^{-1} (\bfrakx - \bfrakb)) ,
   \qquad f \in \kc .
\label{comp-quat-quat-rep}
\en
The {\em Duflo-Moore operator $C$}  is given in the
Fourier domain as the multiplication operator
\be
   (\widehat{C}\widehat{f})(\bfrakk ) = \mathcal C (\bfrakk )
                     \widehat{f} (\bfrakk), \quad \text{where} \quad
   \mathcal C (\bfrakk ) = \left[\frac {2\pi}{\vert \bfrakk\vert}  \right]^2 .
\label{quat-duflo-moore-op}
\en
The admissibility condition is now
$$  (2\pi)^4\int_{\mathbb R^4} \frac {\vert \widehat{f}(\bfrakk )\vert^2}
            {\vert \bfrakk \vert^4} \; d\bfrakk < \infty, $$
and for any two vectors $\eta_1, \eta_2$ in the domain of $C$ and  arbitrary
$f_1, f_2 \in \kc$, the orthogonality relation becomes
\be
  \int_{G^{\mathbb H}_{\text{aff}}} \langle f_1 \mid
  U_{\mathbb C} (\bfrakb, \bfraka)\eta_1
  \rangle\langle\eta_2 \mid  U_{\mathbb C} (\bfrakb , \bfraka)^*f_2\rangle \;
  d\mu_\ell (\bfrakb , \bfraka)
  = \langle C\eta_2 \mid C\eta_1 \rangle \langle f_1 \mid f_2 \rangle ,
\label{quat-comp-orthog}
\en
with its operator version
\be
  \int_{G^{\mathbb H}_{\text{aff}}}  \vert U_{\mathbb C} (\bfrakb , \bfraka)\eta_1
  \rangle\langle\eta_2 \vert  U_{\mathbb C} (\bfrakb , \bfraka)^* \; d\mu_\ell (\bfrakb , \bfraka)
  = \langle C\eta_2 \mid C\eta_1 \rangle  I_\kc \; .
\label{op-quat-comp-orthog}
\en
Similarly, for $\langle C\eta_2 \mid C\eta_1 \rangle \neq 0$,
\be
  \frac 1{\langle C\eta_2 \mid C\eta_1 \rangle} \int_{G^{\mathbb H}_{\text{aff}}}
      U_{\mathbb C} (\bfrakb , \bfraka)\vert\eta_1 \rangle
      \langle \eta_2 \vert U_{\mathbb C} (\bfrakb , \bfraka)^* \;
      d\mu_\ell (\bfrakb , \bfraka)  =
      I_\kc\; .
\label{quat-comp-resolid}
\en
The family of {\em coherent states or wavelets} are
\be
  \mathfrak S_{\mathbb C} = \{\eta_{\mathfrak b , \mathfrak a} =
  U_{\mathbb C}(\bfrakb , \bfraka )\eta \mid
     (\bfrakb , \bfraka ) \in {G^{\mathbb H}_{\text{aff}}}\} ,
\label{quat-cs-wav}
\en
with the resolution of the identity,
\be
   \int_{G^{\mathbb H}_{\text{aff}}}
      \vert\eta_{\mathfrak b, \mathfrak a} \rangle
      \langle \eta_{\mathfrak b , \mathfrak a} \vert \;  d\mu_\ell (\bfrakb , \bfraka)  =
      I_\kc \; .
\label{quat-comp-wav-resolid}
\en

\subsection{UIR of $G^{\mathbb K}_{\text{aff}}$ in a complex Hilbert space}\label{subsec-gen-uir-compl}
From the above and the well-known representations of $G^{ \mathbb R}_{\text{aff}}$
and $G^{ \mathbb C}_{\text{aff}}$, we may write down a general expression for the UIR of
$G^{ \mathbb K}_{\text{aff}}$ for any three of the values of $\mathbb K$.


\section{UIR of $G^{\mathbb H}_{\text{aff}}$ in a quaternionic Hilbert space}\label{sec-uir-quat-sp}
We now proceed to construct a unitay irreducible representation of the quaternionic affine group
$G^{\mathbb H}_{\text{aff}}$  on a quaternionic Hilbert space. It will turn out that this representation has an intimate connection with the representation
$U_{\mathbb C}(\bfrakb , \bfraka)$ in (\ref{comp-quat-quat-rep}) on $\kc$.

\subsection{A right quaternionic Hilbert space}\label{subsec-quat-hilb-sp}
We consider the Hilbert space $\hquat$, of quaternion valued functions over the quaternions.
An element  $\bfrakf \in \hquat$ has the form
\be
 \bfrakf (\bfrakx) = \begin{pmatrix} f_1 (\bfrakx ) & -\overline{f_2 (\bfrakx )}\\
                                      f_2 (\bfrakx ) & \overline{f_1 (\bfrakx )}\end{pmatrix},
                                      \quad \bfrakx \in \mathbb H ,
\label{hquat-vect}
\en
where $f_1$ and $f_2$ are two complex valued functions. The norm in $\hquat$ is given by
\be
  \Vert \bfrakf\Vert_{\mathfrak H_\quat}^2 = \int_{\mathbb H}\bfrakf (\bfrakx )^\dag \bfrakf (\bfrakx )
  \; d\bfrakx = \int_{\mathbb H}\vert\bfrakf (\bfrakx )\vert^2 \; d\bfrakx
      = \left[\int_{\quat} \left(\;\vert f_1 (\bfrakx ) \vert^2 + \vert f_2 (\bfrakx ) \vert^2\;\right)\; d\bfrakx \;
        \right]\sigma_0 ,
\label{bquat-norm}
\en
the finiteness of which implies that both $f_1$ and $f_2$ have to be elements of $\kc =
L^2_{\mathbb C}(\mathbb H , d\bfrakx )$, so that we may write
$$
   \Vert \bfrakf\Vert_{\mathfrak H_\quat}^2 = \left(\; \Vert f_1 \Vert_{\mathfrak H_\mathbb C}^2 +  \Vert f_2 \Vert_{\mathfrak H_\mathbb C}^2 \; \right)\sigma_0 . $$
In view of this, we may also write $\hquat = L^2_\mathbb H ( \quat , d\bfrakx )$.
In using the ``bra-ket'' notation we shall use the notation and convention:
\be
 \bbra \bfrakf \bmid =   \begin{pmatrix}\langle  f_1 \vert   & \langle f_2 \vert\\
                                     -\langle \overline{f}_2 \vert &  \langle \overline{f}_1 \vert\end{pmatrix},
 \quad \text{and} \quad
  \bmid \bfrakf \bket =   \begin{pmatrix}\vert f_1 \rangle   & -  \vert \overline{f}_2 \rangle\\
                                        \vert f_2 \rangle &  \vert \overline{f}_1 \rangle
                                        \end{pmatrix},
  \label{bra-ket}
 \en
 The scalar product of two vectors $\bfrakf, \bfrakf' \in \hquat$ is
 \bea
  \bbra \bfrakf \bmid \bfrakf' \bket & = & \int_{\mathbb H}\bfrakf (\bfrakx )^\dag \bfrakf' (\bfrakx )
  \; d\bfrakx \nonumber\\
  & = & \begin{pmatrix}
  \langle f_1 \mid f_1^\prime\rangle_{\mathfrak H_\mathbb C} +
  \langle f_2 \mid f_2^\prime\rangle_{\mathfrak H_\mathbb C}
 & -\langle f_2^\prime \mid \overline{f}_1\rangle_{\mathfrak H_\mathbb C} +
 \langle f_1^\prime \mid \overline{f}_2\rangle_{\mathfrak H_\mathbb C} \\
  \langle \overline{f}_2^\prime \mid {f}_1\rangle_{\mathfrak H_\mathbb C} -
 \langle \overline{f}_1^\prime \mid {f}_2\rangle_{\mathfrak H_\mathbb C}
 & \langle f_1^\prime \mid f_1 \rangle_{\mathfrak H_\mathbb C} +
  \langle f_2^\prime \mid f_2\rangle_{\mathfrak H_\mathbb C} .
\end{pmatrix}
\label{hquat-sc-prod}
  \ena
  Note that
$$ \bbra \bfrakf \bmid \bfrakf' \bket^\dag =\bbra \bfrakf' \bmid \bfrakf \bket . $$
We see that if $\bfrakf$ is orthogonal to $\bfrakf'$ in $\hquat$ then
$$ \langle f_1 \mid f_1^\prime\rangle_{\mathfrak H_\mathbb C} +
  \langle f_2 \mid f_2^\prime\rangle_{\mathfrak H_\mathbb C} =
    \langle \overline{f}_2^\prime \mid {f}_1\rangle_{\mathfrak H_\mathbb C} -
 \langle \overline{f}_1^\prime \mid {f}_2\rangle_{\mathfrak H_\mathbb C}  = 0 .$$
In other words, defining the two vectors $\mathbf f = \begin{pmatrix} f_1 \\ f_2 \end{pmatrix} , \; \mathbf f' = \begin{pmatrix} f_1'\\ f_2'\end{pmatrix} \in
\kc \oplus \kc$, the orthogonality of  $\bfrakf$ and  $\bfrakf'$ in $\hquat$ implies
$\langle \mathbf f \mid \mathbf f' \rangle = 0$, i.e., the  orthogonality of   $\mathbf f$
and $\mathbf f'$
in $\kc \oplus \kc$ and in addition  (in an obvious notation), that  $\mathbf f \wedge \mathbf f' = 0$.

Multiplication by quaternions on $\hquat$ is defined from the right:
$$
  (\hquat \times \quat )\ni (\bfrakf , \bfrakq )\longmapsto \bfrakf \bfrakq, \quad \text{such that}
  \quad (\bfrakf \bfrakq) (\bfrakx ) = \bfrakf (\bfrakx )\bfrakq , $$
i.e., we take $\hquat$ to be a right quaternionic Hilbert space. This convention is consistent with the scalar product (\ref{hquat-sc-prod}) in the sense that
$$ \bbra \bfrakf \bmid \bfrakf'\bfrakq \bket = \bbra \bfrakf \bmid \bfrakf' \bket\bfrakq  \quad
\text{and} \quad \bbra \bfrakf\bfrakq \bmid \bfrakf' \bket =
\bfrakq^{\dagger}\bbra \bfrakf \bmid \bfrakf' \bket .$$

 On the other hand,
the action of operators $\bA$ on vectors $\bfrakf \in \hquat$ will be from the left $(\bA, \bfrakf)
\longmapsto \bA\bfrakf$. In particular, an operator $A$ on $\kc$ defines an operator $\bA$ on $\hquat$
as,
$$
   (\bA\bfrakf)(\bfrakx) = \begin{pmatrix} (Af_1) (\bfrakx ) & -\overline{(Af_2) (\bfrakx )}\\
                                      (Af_2) (\bfrakx ) & \overline{(Af_1) (\bfrakx )}\end{pmatrix} .$$
Multiplication of operators by quaternions will also be from the left. Thus, $\bfrakq\bA$ acts on
the vector $\bfrakf$ in the manner
$$ (\bfrakq\bA\bfrakf)(\bfrakx) = \bfrakq (\bA\bfrakf)(\bfrakx) . $$
We shall also need the``rank-one operator''
\bea
\bmid \bfrakf \bket\bbra \bfrakf' \bmid  & = &
\begin{pmatrix} \vert f_1 \rangle& - \vert \overline{f}_2 \rangle\\
                                        \vert f_2 \rangle &  \vert \overline{f}_1 \rangle
                                        \end{pmatrix}
 \begin{pmatrix}\langle  f_1' \vert  & \langle f_2' \vert\\
              -\langle \overline{f}_2' \vert &  \langle \overline{f}_1' \vert\end{pmatrix}
              \nonumber \\
    & = & \begin{pmatrix} \vert f_1 \rangle\langle f_1'\vert
                           + \vert \overline{f}_2  \rangle\langle \overline{f}_2'\vert
               & \vert f_1 \rangle\langle f_2'\vert
                           - \vert \overline{f}_2  \rangle\langle \overline{f}_1'\vert \\
                  -\vert \overline{f}_1 \rangle\langle \overline{f}_2'\vert
                           + \vert {f}_2  \rangle\langle {f}_1'\vert
               &  \vert \overline{f}_1  \rangle\langle \overline{f}_1'\vert
                          + \vert f_2 \rangle\langle f_2'\vert\; .
 \end{pmatrix}
 \label{rank-one-op}
\ena

An orthonormal basis in $\hquat$ can be  built using an orthonormal basis in $\kc$. Indeed, let
$\{\phi_n\}_{n=0}^\infty$ be an orthonormal basis of $\kc = L^2_{\mathbb C} (\quat, d\bfrakx )$. Define the
vectors
\be
  \bmid \bPhi_n\bket =  \frac 1{\sqrt{2}} \begin{pmatrix} \vert \phi_n \rangle&  \vert \phi_n \rangle\\
                                        -\vert \overline{\phi}_n \rangle &  \vert \overline{\phi}_n \rangle
                                        \end{pmatrix}, \quad n = 0,1,2, \ldots ,
\label{quat-onb}
\en
in $\hquat$. It is easy to check that these vectors are orthonormal in $\hquat$. The fact that they form a basis follows from the fact that the vectors $\{\phi_n\}_{n=0}^\infty$  are a basis of
$L^2_{\mathbb C} (\quat, d\bfrakx )$. Indeed, with
$$
      \bmid \bfrakf \bket = \begin{pmatrix} \vert f_1 \rangle& - \vert \overline{f}_2 \rangle\\
                                        \vert f_2 \rangle &  \vert \overline{f}_1 \rangle
                                        \end{pmatrix} \in  L^2_{\mathbb H} (\quat, d\bfrakx ), $$
and writing
$$
 \vert f_1 \rangle = \sum_{n=0}^\infty b_n \vert\phi_n\rangle, \;\;
 \vert f_2 \rangle = \sum_{n=0}^\infty c_n \vert\phi_n\rangle,  \quad \text{with} \quad
  b_n = \langle \phi_n \mid f_1\rangle ,\;\;  c_n = \langle \phi_n \mid f_2\rangle , $$
we easily verify that
$$  \bmid \bfrakf \bket = \sum_{n=0}^\infty\bmid \bPhi_n\bket \bfrakq_n , $$
where
$$ \bfrakq_n = \bbra \bPhi_n \bmid \bfrakf\bket =
\frac 1{\sqrt{2}}\begin{pmatrix}
  \langle \phi_n \mid f_1\rangle_{\mathfrak H_\mathbb C} -
  \langle \overline{\phi}_n \mid f_2\rangle_{\mathfrak H_\mathbb C}
 & -\langle f_2 \mid \overline{\phi}_n\rangle_{\mathfrak H_\mathbb C} -
 \langle f_1 \mid \phi_n\rangle_{\mathfrak H_\mathbb C} \\
  \langle \overline{f}_2 \mid \phi_n\rangle_{\mathfrak H_\mathbb C} +
 \langle \overline{f}_1 \mid \overline{\phi}_n\rangle_{\mathfrak H_\mathbb C}
 & \langle f_1 \mid \phi_n \rangle_{\mathfrak H_\mathbb C} -
  \langle f_2 \mid \overline{\phi}_n\rangle_{\mathfrak H_\mathbb C}
\end{pmatrix} .$$
In verifying the above, one needs to take into account the fact that the vectors
$\{\overline{\phi}_n\}_{n = 0}^\infty$ also form an orthonormal basis of $\kc$.

\subsection{Representation of $G^{\mathbb H}_{\text{aff}}$  on $\hquat$}\label{subsec-the-rep}
A representation of $G^{\mathbb H}_{\text{aff}}$  on $\hquat$ can be obtained by simply
transcribing (\ref{comp-quat-quat-rep}) into the present context. We define the operators
$\bU_\quat (\bfrakb, \bfraka )$ on $\hquat$:
\be
  (\bU_\quat (\bfrakb, \bfraka )\bfrakf )(\bfrakx) = \frac 1{\text{det}[\bfraka]}
   \bfrakf( \bfraka^{-1} (\bfrakx - \bfrakb)) ,
   \qquad \bfrakf \in \hquat ,
\label{quat-quat-quat-rep}
\en
which by (\ref{comp-quat-quat-rep}) and (\ref{bra-ket}) can also be written as
\be
  \bmid \bU_\quat (\bfrakb, \bfraka )\bfrakf \bket =
            \begin{pmatrix}\vert U_\mathbb C (\bfrakb, \bfraka )f_1 \rangle   & -  \vert \overline{U_\mathbb C (\bfrakb, \bfraka )f}_2 \rangle\\
 \vert U_\mathbb C (\bfrakb, \bfraka ) f_2 \rangle &
  \vert \overline{U_\mathbb C (\bfrakb, \bfraka )f}_1 \rangle
                                        \end{pmatrix}.
\label{quat-quat-quat-rep2}
\en
The unitarity of this representation is easy to verify. Indeed,
\beano
 \Vert\bU_\quat (\bfrakb, \bfraka )\bfrakf\Vert^2  & =  &
  \int_\quat \vert(\bU_\quat (\bfrakb, \bfraka )\bfrakf )(\bfrakx)\vert^2\; d\bfrakx \\
 & = & \int_\quat \left(\vert (U_\mathbb C (\bfrakb, \bfraka )f_1) (\bfrakx)\vert^2 +
  \vert (U_\mathbb C (\bfrakb, \bfraka )f_2 )(\bfrakx)\vert^2\right)\; d\bfrakx \; \sigma_0 ,
\enano
which, by the unitarity of   the representation $U_\mathbb C (\bfrakb, \bfraka )$ on $\kc$ gives
$$
 \Vert\bU_\quat (\bfrakb, \bfraka )\bfrakf\Vert^2_{\mathfrak H_\mathbb H} =
 \left(\Vert f_1\Vert^2_{\mathfrak K_\mathbb C} +
  \Vert f_2\Vert^2_{\mathfrak K_\mathbb C}\right) \sigma_0
  = \Vert \bfrakf\Vert^2_{\mathfrak H_\mathbb H} .
$$

Similarly, the irreducibility of $U_\mathbb C (\bfrakb, \bfraka )$ on $\kc$ leads to the
irreducibility of $\bU_\quat (\bfrakb, \bfraka )$. But it will be useful to first prove the
{\em square-integrability} of this representation.

\subsection{Square-integrability of $\bU_\mathbb H (\bfrakb, \bfraka )$}\label{subsec-square-int}
Using the Duflo-Moore operator $C$ in (\ref{quat-duflo-moore-op}) for the representation
$U_{\mathbb C}(\bfrakb , \bfraka)$ (see (\ref{comp-quat-quat-rep}), we define the Duflo-Moore operator $\mathbf C$ for the representation $\bU_\quat (\bfrakb, \bfraka )$:
$$
   (\mathbf C \bfrakf)(\bfrakx)  =  \begin{pmatrix} (Cf_1)(\bfrakx) &
                                     -\overline{(Cf_2) (\bfrakx )}\\
     (Cf_2) (\bfrakx ) & \overline{(Cf_1) (\bfrakx )}\end{pmatrix} .$$
We say that the vector $\bfrakf$ is {\em admissible for the representation
$\bU_\mathbb H (\bfrakb, \bfraka )$} if it is in the domain of $\mathbf C$, i.e., if both $f_1$ and
$f_2$ are admissible for the representation $U_\mathbb C (\bfrakb, \bfraka )$. It is then easy to
see that the set of admissible vectors is dense in $\hquat$.

  Let $\bfrakf$ and $\bfrakf'$ be two admissible vectors. Then from (\ref{quat-quat-quat-rep2}),
(\ref{rank-one-op}) and (\ref{op-quat-comp-orthog}) we get
\be
   \int_{G^\mathbb H_{\text{aff}}}\bmid \bU_\quat (\bfrakb, \bfraka )\bfrakf \bket \bbra  \bU_\quat (\bfrakb, \bfraka )\bfrakf' \bmid\!
 d\mu_\ell (\bfrakb , \bfraka) = \bfrakq\; I_{\mathfrak H_\mathbb H},
\label{quat-sq-int-cond}
\en
where $\bfrakq$ denotes the operator of multiplication from the left, on the Hilbert space $\hquat$,
by the quaternion
\be
\bfrakq =
\begin{pmatrix}
  \langle Cf'_1 \mid Cf_1\rangle_{\mathfrak H_\mathbb C} +
  \langle \overline{Cf'}_2 \mid \overline{Cf}_2\rangle_{\mathfrak H_\mathbb C}
 & \langle Cf'_2 \mid Cf_1\rangle_{\mathfrak H_\mathbb C} -
 \langle \overline{Cf'}_1 \mid \overline{Cf}_2\rangle_{\mathfrak H_\mathbb C} \\
  \langle Cf'_1 \mid Cf_2\rangle_{\mathfrak H_\mathbb C} -
 \langle \overline{Cf'}_2 \mid \overline{Cf}_1\rangle_{\mathfrak H_\mathbb C}
 & \langle \overline{Cf'}_1 \mid \overline{Cf}_1 \rangle_{\mathfrak H_\mathbb C} +
  \langle Cf'_2 \mid Cf_2\rangle_{\mathfrak H_\mathbb C}
\end{pmatrix} .
\label{const-mat}
\en
Note that defining  transposed vectors
$$
  \bmid \bfrakf^T\bket =\begin{pmatrix} \vert f_1 \rangle &  \vert f_2 \rangle\\
                                        -\vert \overline{f}_2 \rangle &  \vert \overline{f}_1 \rangle
                                        \end{pmatrix},
                                        $$
we get,
$$
\bfrakq = \bbra \mathbf C\bfrakf^{\;\prime T} \bmid \mathbf C\bfrakf^T \bket^T . $$
Equation (\ref{quat-sq-int-cond}) expresses the square-integrability condition for the representation
 $\bU_\mathbb H (\bfrakb, \bfraka )$. In particular, with $\bfrakf = \bfrakf'$, we get the resolution
 of the identity,
\be
 \left[ \Vert \mathbf C\bfrakf\Vert^2_{\mathfrak H_{\mathbb H}}\right]^{- 1}
 \int_{G^\mathbb H_{\text{aff}}}\bmid \bU_\quat (\bfrakb, \bfraka )\bfrakf \bket \bbra  \bU_\quat (\bfrakb, \bfraka )\bfrakf \bmid
 d\mu_\ell (\bfrakb , \bfraka) = I_{\mathfrak H_\mathbb H}.
\label{quqt-quat-resoln-id}
\en

\subsection{Irreducibility of $\bU_\mathbb H (\bfrakb, \bfraka )$ }\label{subsec-irreducibility}
 Let $\bfrakf$ be any non-zero vector
in $\hquat$ and let $\bfrakf'$ be any vector such that
$\bbra\bU_\quat (\bfrakb, \bfraka )\bfrakf\bmid \bfrakf'\bket = 0$ for all
$(\bfrakb, \bfraka ) \in G^{\mathbb H}_{\text{aff}}$. We will show that this implies
$\bfrakf' =0$, in other words every vector in $\hquat$ is cyclic for the representation
$\bU_\mathbb H (\bfrakb, \bfraka )$  and hence it is irreducible. From (\ref{quqt-quat-resoln-id})
we see that every admissible vector is cyclic for the representation $\bU_\mathbb H (\bfrakb, \bfraka )$. On the other hand, the set of admissible vectors is dense in $\hquat$. From this it follows that
every vector in $\hquat$ is cyclic.

\subsection{Wavelets and reproducing kernels}\label{sec-wav-rep-ker}
Let $\eta \in \hquat$ be an addmissible vector for the representation
$\bU_\mathbb H (\bfrakb, \bfraka )$, normalized so that
$$   \Vert\mathbf C\eta\Vert^2 = 1 . $$
We define the {\em quaternionic wavelets or coherent states\/} to be the vectors
\be
  \mathfrak S_{\mathbb H} = \{\bfeta_{\mathfrak b , \mathfrak a} =
  \bU_{\mathbb H}(\bfrakb , \bfraka )\eta \mid
     (\bfrakb , \bfraka ) \in {G^{\mathbb H}_{\text{aff}}}\} ,
\label{quat-quat-cs-wav}
\en
in $\hquat$.
By virtue of (\ref{quat-sq-int-cond}) they satisfy the resolution of the identity
\be
   \int_{G^{\mathbb H}_{\text{aff}}}
      \vert\bfeta_{\mathfrak b, \mathfrak a} \bket
      \bbra \bfeta_{\mathfrak b , \mathfrak a} \vert \;  d\mu_\ell (\bfrakb , \bfraka)  =
      I_{\mathfrak H_\mathbb H} \; .
\label{quat-quat-wav-resolid}
\en
There is the associated reproducing kernel $\bK : G^{\mathbb H}_{\text{aff}} \times
G^{\mathbb H}_{\text{aff}} \longrightarrow \quat$,
\be
  \bK (\overline{\bfrakb} , \overline{\bfraka} ; \; \bfrakb' , \bfraka') = \bbra
  \bfeta_{\mathfrak b, \mathfrak a} \bmid \bfeta_{\mathfrak b^\prime, \mathfrak a^\prime}
  \bket_{\mathfrak H_\mathbb H}\; ,
\label{quat-repker}
\en
with the usual properties,
\bea
 \bK (\overline{\bfrakb} , \overline{\bfraka} ; \; \bfrakb' , \bfraka')
 = \overline{\bK (\overline{\bfrakb'} , \overline{\bfraka'} ; \; \bfrakb , \bfraka)},
 & \qquad & \bK (\overline{\bfrakb} , \overline{\bfraka} ; \; \bfrakb , \bfraka)
 > 0, \nonumber\\
 \int_{G^{\mathbb H}_{\text{aff}}} \bK (\overline{\bfrakb} , \overline{\bfraka} ;
 \; \bfrakb'' , \bfraka'')\;
 \bK (\overline{\bfrakb''} , \overline{\bfraka''} ; \; \bfrakb' , \bfraka') \;d\mu_\ell
 (\bfrakb'' , \bfraka'') & = &
 \bK (\overline{\bfrakb} , \overline{\bfraka} ; \; \bfrakb' , \bfraka'),
\ena
(recall that in our representation of quaternions, $\overline{\bfrakq} = \bfrakq^\dag$).

\section{conclusion}
We have constructed the CWT of the quaternionic affine group on a complex and on a quaternionic Hilbert space along the lines of its real and complex counterparts. The resulting resolution of the identity and thereby a reconstruction formula has also been obtained.  It may be possible to extend the UIR $U_{\HQ}(\bfrakb,\bfraka)$ to a larger class by multiplying $U_{\HQ}(\bfrakb,\bfraka)$ from the right by the $SU(2)$ part of $\bfraka$. Further, one may attempt to discretize the CWT which may enlarge the scope of applications. It may also be interesting to look at the quaternionic case from the multiresolution analytic point of view. We intend to look at some of these issues  in the future.

\end{document}